\documentclass[12pt,preprint]{aastex}
\usepackage{graphicx}

\shorttitle{Non-detection of HD 97658b transits}

\shortauthors{Dragomir et al.}

\begin{document}

\renewcommand\thefootnote{\fnsymbol{footnote}} 

\title{Non-detection of previously reported transits of HD 97658\lowercase{b} with MOST\footnote{Based on data from the MOST satellite, a Canadian Space Agency mission operated by Microsatellite Systems Canada Inc. (MSCI; former Dynacon Inc.) and the Universities of Toronto and British Columbia, with the assistance of the University of Vienna.} photometry}

\renewcommand\thefootnote{\fnsymbol{footnote}}

\author{
  Diana Dragomir\altaffilmark{1},
  Jaymie M. Matthews\altaffilmark{1},
  Andrew W. Howard\altaffilmark{2,3},
  Victoria Antoci\altaffilmark{1,4,5}, 
  Gregory W. Henry\altaffilmark{6},
  David B. Guenther\altaffilmark{7},
  John A. Johnson\altaffilmark{8,9},
  Rainer Kuschnig\altaffilmark{5},
  Geoffrey W. Marcy\altaffilmark{2},
  Anthony F. J. Moffat\altaffilmark{10},
  Jason F. Rowe\altaffilmark{11},
  Slavek M. Rucinski\altaffilmark{12},
  Dimitar Sasselov\altaffilmark{13}, 
  Werner W. Weiss\altaffilmark{5}
}

\email{diana@phas.ubc.ca}
\altaffiltext{1}{Department of Physics and Astronomy, University of British Columbia, Vancouver, BC V6T1Z1, Canada}
\altaffiltext{2}{Department of Astronomy, University of California, Berkeley, CA 94720-3411, USA}
\altaffiltext{3}{Space Sciences Laboratory, University of California, Berkeley, CA 94720-7450, USA}
\altaffiltext{4}{Stellar Astrophysics Centre (SAC), Department of Physics and Astronomy, Aarhus University, Ny Munkegade 120, DK-8000 Aarhus C, Denmark}
\altaffiltext{5}{Universit\"{a}t Wien, Institut f\"{u}r Astronomie, T\"{u}rkenschanzstrasse 17, AÐ1180 Wien, Austria}
\altaffiltext{6}{Center of Excellence in Information Systems, Tennessee State University, 3500 John A. Merritt Blvd., Box 9501, Nashville, TN 37209, USA}
\altaffiltext{7}{Department of Astronomy and Physics, St. MaryÕs University Halifax, NS B3H 3C3, Canada}
\altaffiltext{8}{Department of Astronomy, California Institute of Technology, 1200 East California Boulevard, MC 249-17, Pasadena, CA 91125, USA}
\altaffiltext{9}{NASA Exoplanet Science Institute (NExScI), CIT Mail Code 100-22, 770 South Wilson Avenue, Pasadena, CA 91125, USA}
\altaffiltext{10}{D\'{e}pt de physique, Univ de Montr\'{e}al C.P. 6128, Succ. Centre-Ville, Montr\'{e}al, QC H3C 3J7, and Obs. du mont M\'{e}gantic, Canada} 
\altaffiltext{11}{NASA Ames Research Center, Moffett Field, CA 94035}
\altaffiltext{12}{Department of Astronomy and Astrophysics, University of Toronto, 50 St. George Street, Toronto, ON M5S 3H4, Canada}
\altaffiltext{13}{Harvard-Smithsonian Center for Astrophysics, 60 Garden Street, Cambridge, MA 02138, USA}


\begin{abstract}

The radial velocity-discovered exoplanet HD 97658b was recently announced to transit, with a derived planetary radius of 2.93 $\pm$ 0.28 R$_{\oplus}$. As a transiting super-Earth orbiting a bright star, this planet would make an attractive candidate for additional observations, including studies of its atmospheric properties. We present and analyze follow-up photometric observations of the HD 97658 system acquired with the MOST space telescope. Our results show no transit with the depth and ephemeris reported in the announcement paper. For the same ephemeris, we rule out transits for a planet with radius larger than 2.09 R$_{\oplus}$, corresponding to the reported 3$\sigma$ lower limit. We also report new radial velocity measurements which continue to support the existence of an exoplanet with a period of 9.5 days, and obtain improved orbital parameters.
\end{abstract}


\section{Introduction}

Known transiting super-Earth exoplanets around relatively bright ($V < 10$) stars are still scarce. The first to be announced was 55 Cnc e (\citealt{Win11}; \citealt{Dem11}), and more recently transits of HD 97658b were reported by \cite{Hen11}. 

HD 97658b was discovered through Keck-HIRES radial velocity (RV) measurements \citep{How11}. This planet orbits a K1 dwarf ($V \sim 7.8$) with a period of 9.494 $\pm$ 0.005 days. Its low discovery minimum mass ($m\sin i$ = $8.2 \pm 1.2$ M$_{\oplus}$; \citealt{How11}) suggests a potential super-Earth. 

The system was observed near mid-transit times predicted by the RV ephemeris with two Automated Photometric Telescopes (APTs; by \citealt{Hen11}, hereafter H11). Four partial transit windows were covered, each two planetary orbital cycles apart (since the orbital period is almost exactly 9.5 days, so if one window occurs at night for a given observing site, the next is during daytime). The individual light curves suggested the presence of transits, and the signal persisted when the photometry was phased with the period predicted from the RV ephemeris. Measurements obtained during nights predicted to be out of transit were consistent with a constant flux model, supporting the apparent transit detection. The feature identified as a transit had a depth of 1470 $\pm$ 260 ppm, corresponding to a 2.93 $\pm$ 0.28 R$_{\oplus}$ planet with an orbital inclination of 90 $\pm$ 1$^{\circ}$. These values would mean a relatively low density (1.40$^{+0.53}_{-0.36}$ g cm$^{-3}$) super-Earth. 

Given the brightness of its host star, a transiting HD 97658b would be an ideal candidate for follow-up observations aiming to improve the accuracy and precision of the parameters of both the planet and its host star. In addition, these characteristics together with a potentially large atmospheric scale height would make it suitable for detailed atmospheric studies within reach of current space-based and large ground-based facilities, which would contribute to the much needed insight into the physical properties of super-Earth exoplanets.

Motivated by these stakes, we acquired space-based photometry of HD 97658 to better characterize the transit. In this Letter we report and analyze this photometry as well as 73 new unpublished Keck RV measurements of the star. Though the RVs continue to confirm the existence of the planet, the results of our photometric analysis show there is no transit with the previously reported depth and  mid-transit times. We note that G. Henry has acquired new APT photometry of the HD 97658 system during the 2012 observing season, and these observations have also failed to confirm the previously reported transits.

\section{MOST Photometry}

\subsection{Observations}

\begin{figure*}[!t]
\includegraphics[scale=0.2]{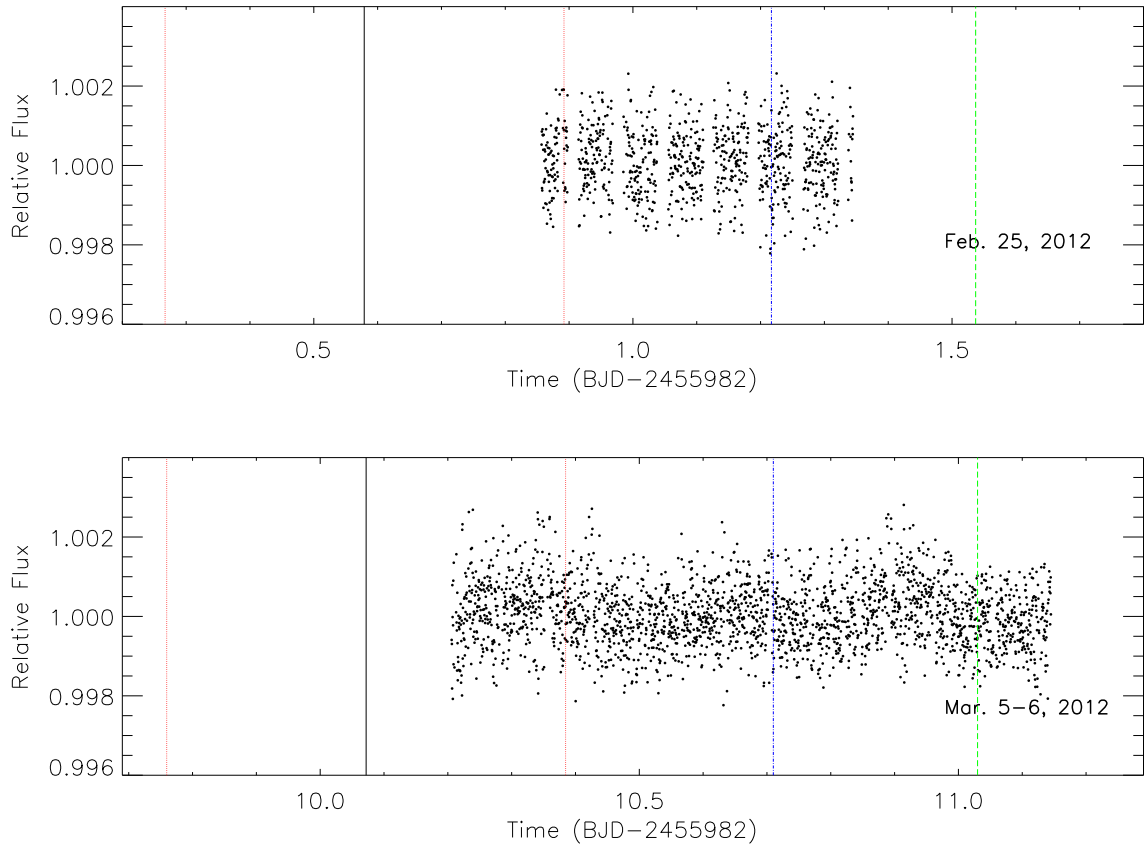}
\caption{Reduced MOST photometry acquired during the predicted transit window of HD 97658b on February 25 ({\it top}) and March 5-6 ({\it bottom}). The gaps in the top light curve correspond to the data removed due to increased stray light (see Section 2.1 in the text for details). The solid black vertical bar corresponds to the predicted mid-transit time based on the ephemeris derived only from the RVs. The red dotted, blue dot-dashed and green dashed vertical bars represent the 1, 2 and 3$\sigma$ limits of the transit window, based on the ephemeris derived only from RVs.}
\end{figure*}

We used the MOST ({\it Microvariability and Oscillations of STars}) space telescope to obtain high-precision photometry of HD 97658. MOST (\citealt{Wal03}, \citealt{Mat04}) is a micro-satellite which carries a 15 cm optical telescope feeding a CCD photometer. It is in a Sun-synchronous polar orbit with a period of 101.4 minutes, and its custom broadband filter covers the optical spectrum from 350 to 700 nm.

HD 97658 was observed contiguously in Direct Imaging mode on February 25, 2012 for 11.8 hours, and again on March 5-6, 2012 for 22.8 hours. The exposure time was 1.5 s, but the observations were stacked on board the satellite in groups of 21 for an integration time of 31.7 s per data point. The raw data was reduced using aperture photometry. The reduction pipeline (described in \citealt{Row08}) corrects for cosmic ray hits and stray light from scattered Earthshine, which varies with the period of the satellite. During the February 25 observations, several sections of the light curve (the gaps in Figure 1) were removed due to high levels of stray light caused by the passage of the satellite through the South Atlantic Anomaly (SAA), resulting in a final time series containing 1157 points. The final March 5-6 data set contains 2638 points. Both light curves are shown at the same scale in Figure 1. 

The first light curve spans nearly the full 3$\sigma$ transit window for HD 97658b predicted using the ephemeris from H11, (see the top panel of Figure 2) which is based on a joint analysis of the RVs and APT photometry. The second light curve covers a significantly longer window as shown in the middle panel of Figure 2, virtually eliminating the possibility that a transit was missed.

In Figure 1, we also show the predicted mid-transit times and boundaries of the 1, 2 and 3$\sigma$ transit window based on the ephemeris obtained from the RVs alone. Our observations only cover up to about 45\% of the RV-only 3$\sigma$ transit window.

\subsection{Analysis}

The top and middle panels of Figure 2 show the binned photometry for the first and second data set. The bottom panel shows both time series phased with the orbital period of HD 97658b, then binned. The bin size was chosen to obtain approximately 20-25 in-transit points. The planetary and stellar radii derived in H11 are 2.93 R$_{\oplus}$ and 0.70$R_{\odot}$, respectively. We used these values and non-linear limb darkening coefficients ($c_{1}=$0.685, $c_{2}=$-0.669, $c_{3}=$1.405, $c_{4}=$-0.567; A. Prsa, private communication) generated for the MOST bandpass assuming the stellar properties described in \cite{How11} ($T_{eff} = 5170 \pm 44$ K; log $g = 4.63 \pm 0.06$; $[Fe/H] = -0.23 \pm 0.03$) to simulate the predicted transit signature in the MOST light curves. The transit models employed are those from \cite{Man02}. We find an expected depth of 1920 ppm for a limb-darkened transit in the MOST bandpass.

To evaluate the sensitivity of the MOST photometry to transits of a planet with the reported transit parameters, we carried out a transit injection and recovery test. We injected artificial limb-darkened transits of three different depths, corresponding to planetary radii of 2.93, 2.65 and 2.09 $R_{\oplus}$, into the raw light curve. These values represent the radius, and the 1$\sigma$ and 3$\sigma$ lower limits on the radius derived by H11, respectively. For each radius, we injected transits at 100 phases randomly distributed within the 3$\sigma$ transit window based on the H11 ephemeris. We then reduced the resulting 300 light curves (each consisting of the two MOST data sets of the HD 97658 system and thus including two artificial transits per light curve) as described in Section 2.1. The recovered transits were shallower by 12.6, 11 and 4\% than the injected transits for the 2.93, 2.65 and 2.09 $R_{\oplus}$ planetary radii, respectively. This suppression is due to the combined effect of the correlated noise present in the light curve and the data reduction procedure. The recovered transit depths corresponding to the 2.93, 2.65 and 2.09 $R_{\oplus}$ planetary radii had 1$\sigma$ scatter of 16, 20 and 30\%, respectively.

These values demonstrate that the precision of the MOST photometry allows us to verify the existence of a transit with the claimed depth, which we would have detected with 6.3$\sigma$ significance. A transit with a depth corresponding to the 3$\sigma$ lower limit on the reported value would have been detected with 3.3$\sigma$ significance. However, even transits as shallow as this limit are clearly absent from the data, as can be seen in Figure 2. Incorporating the transit depth suppression described above, the solid, dashed and dot-dashed lines correspond to the transit of a planet with radius of 2.93 (claimed value), 2.65 (1$\sigma$ lower limit) and 2.09 R$_{\oplus}$ (3$\sigma$ lower limit), respectively. 

\begin{figure}[!h]
\begin{center}
\includegraphics[scale=0.18]{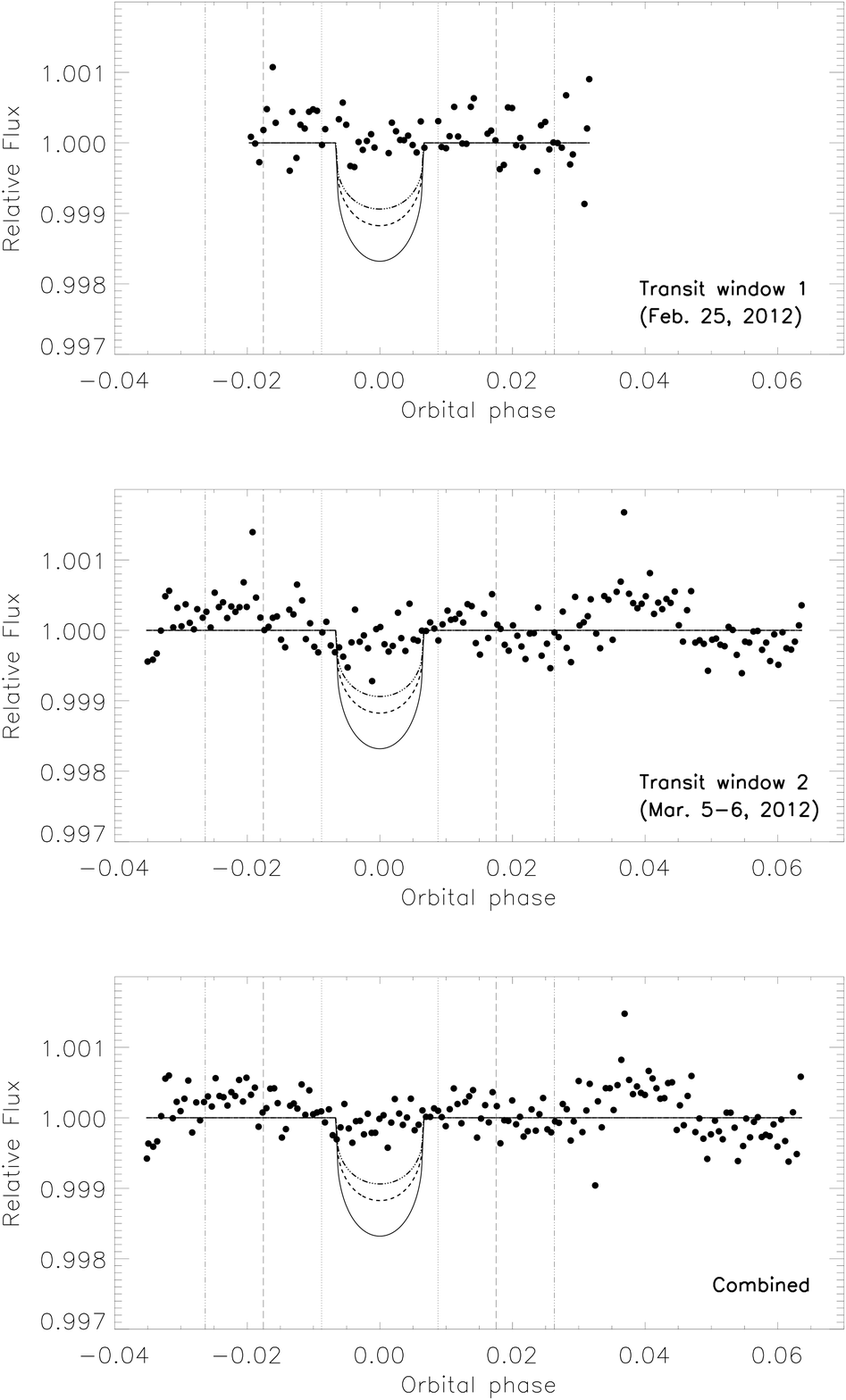}
\caption{HD 97658 MOST photometry acquired on February 25 ({\it top}), on March 5-6 ({\it middle}) and the combined data set ({\it bottom}) phased to the orbital period of the planet and averaged into 9 min phase bins. Transit models with edge-on configuration for a planet with $R_{P}=$2.93 R$_{\oplus}$ (solid line), 2.65 R$_{\oplus}$ (dashed line) and 2.09 R$_{\oplus}$ (dot-dashed line), incorporating the transit depth suppression described in Section 2.2, are shown. The vertical dotted, dashed and dot-dashed lines enclose the 1$\sigma$, 2$\sigma$ and 3$\sigma$ transit windows (based on the H11 ephemeris, which is derived from the RVs and the APT photometry), respectively.}
\end{center}
\end{figure}

We note that our photometry is insufficient for a complete transit search of HD 97658b, since our observations cover less than half the RV-only 3$\sigma$ transit window, as already mentioned in section 2.1.

\section{Keck-HIRES Radial Velocity Measurements}

Since the publication of \cite{How11}, we have measured the radial velocity of HD 97658 for nearly two additional observing seasons (2010--2012) using Keck-HIRES. We used the same observing setup and modeling techniques as in \cite{How11}. The full set of 169 RVs and $S_{\mathrm{HK}}$ measurements of the \ion{Ca}{2} H \& K lines \citep{Isa10} from 2005--2012 are listed in Table \ref{tab:rvs}. The $S_{\mathrm{HK}}$ measurements (indicators of stellar activity) are not correlated with the RVs, consistent with a planetary origin for the RV variation. Seventy-three measurements were recorded between late 2010 and early 2012 after the submission of \cite{How11}, twenty-one of which were acquired after the submission of H11. The RV measurements in common differ slightly here because of modest re-weighting of spectral segments in the Doppler analysis.  

\begin{figure}[!h]
\begin{center}
\includegraphics[scale=0.36]{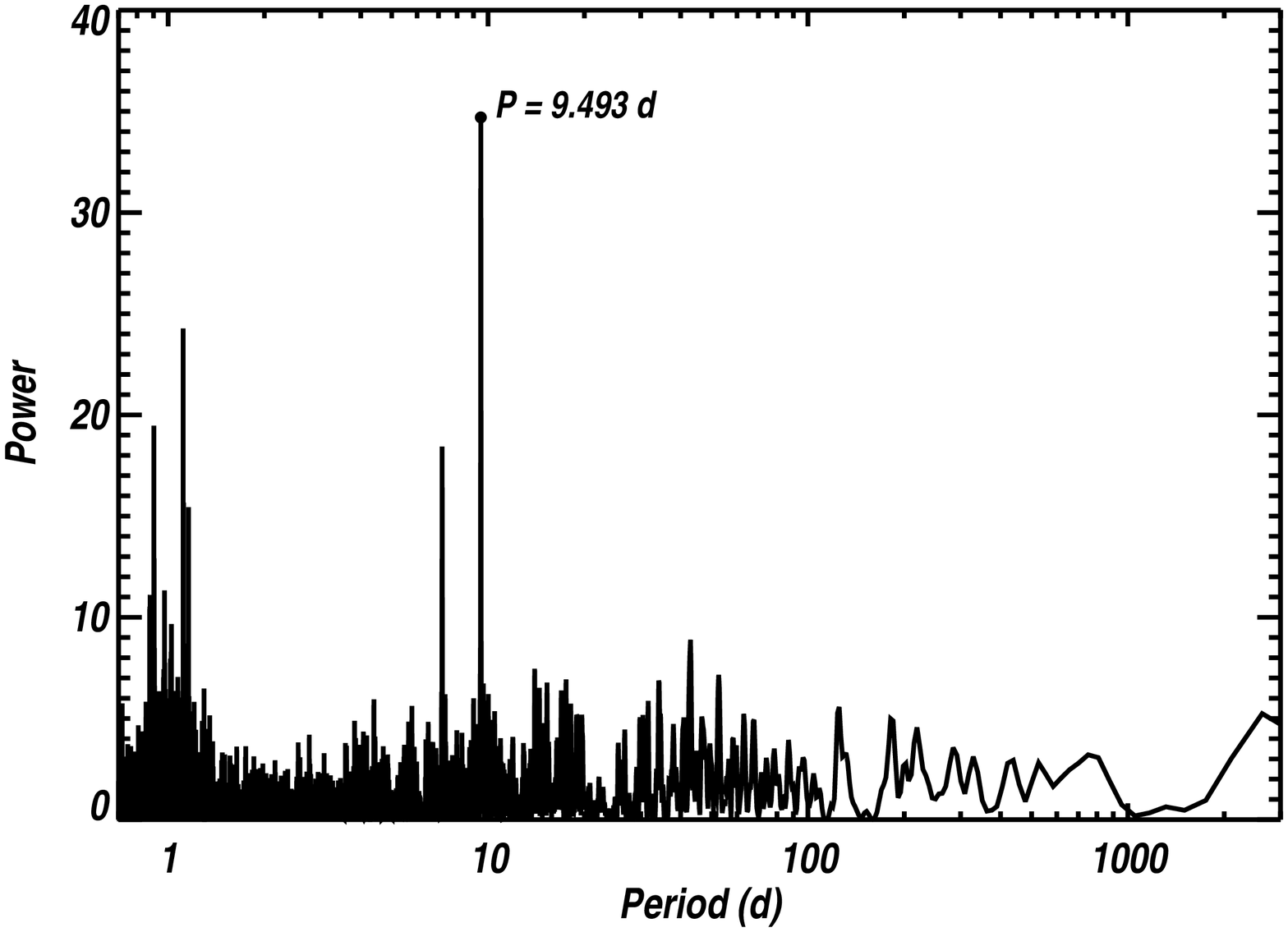}
\includegraphics[scale=0.36]{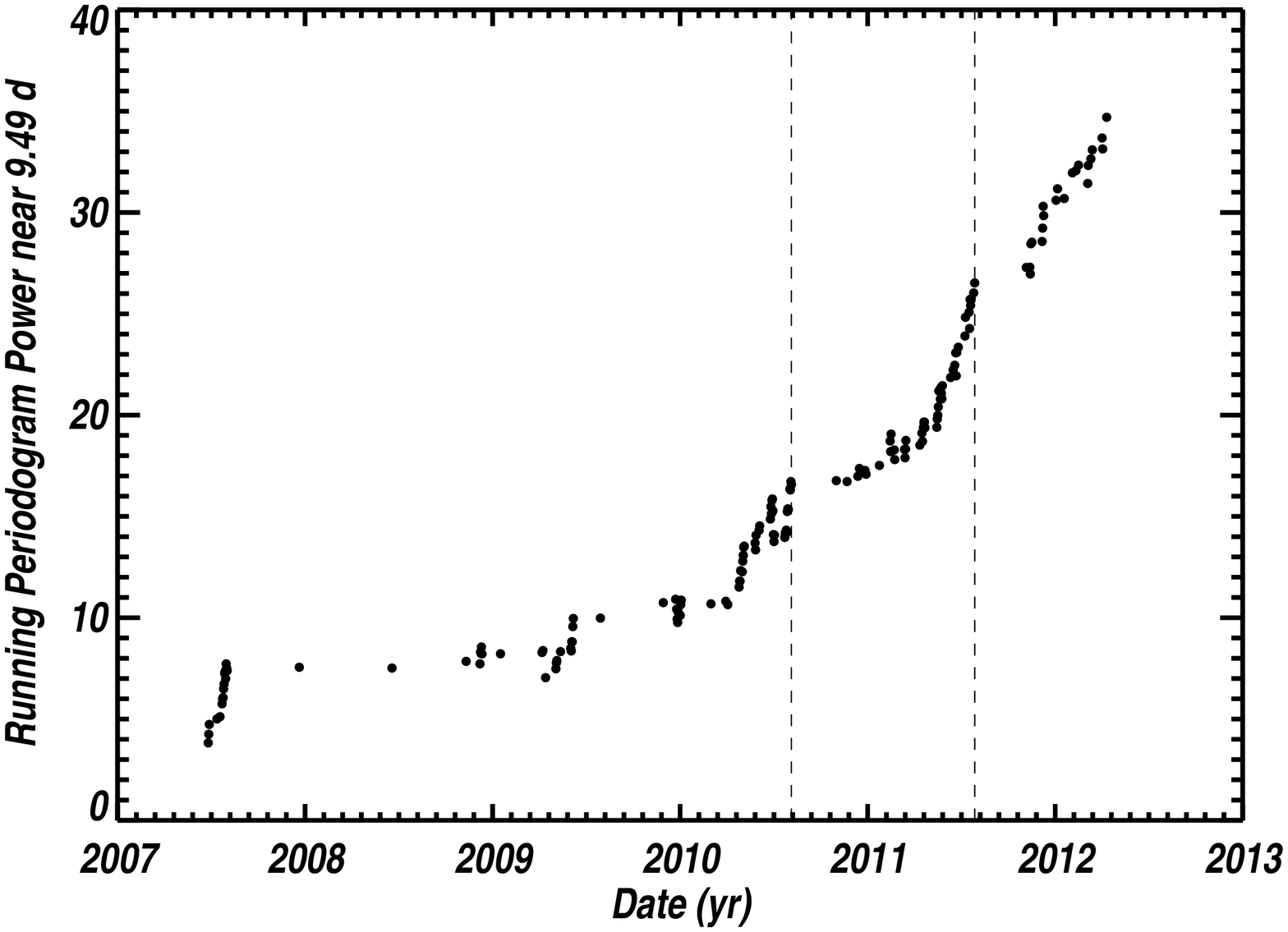}
\caption{{\it Top}: Lomb-Scargle periodogram of all RV measurements of HD 97658. The tall peak near $P=9.493$\,d suggests a planet with that orbital period. {\it Bottom}: Running periodogram power near the orbital period of $P \approx 9.493$ days as a function of time. Each dot represents the maximum periodogram power near 9.493 days for the set of RVs up to that point in time. Only periods within 1\% of the best-fit orbital period of HD 97658 b were used to compute each maximum periodogram power value. The nearly monotonic rise confirms that the periodic RV signal was present throughout the observations and points to a dynamical origin, supporting the existence of the planet. The dashed lines at 2010.7 and 2011.6 correspond to the cut-off dates for the RVs included in \cite{How11} and H11, respectively. The RVs gathered after H11 continue to provide additional support for the planet.}
\label{fig:running_periodogram}
\end{center}
\end{figure}

\begin{figure}[!t]
\begin{center}
\vspace*{0.1in}
\includegraphics[scale=0.9]{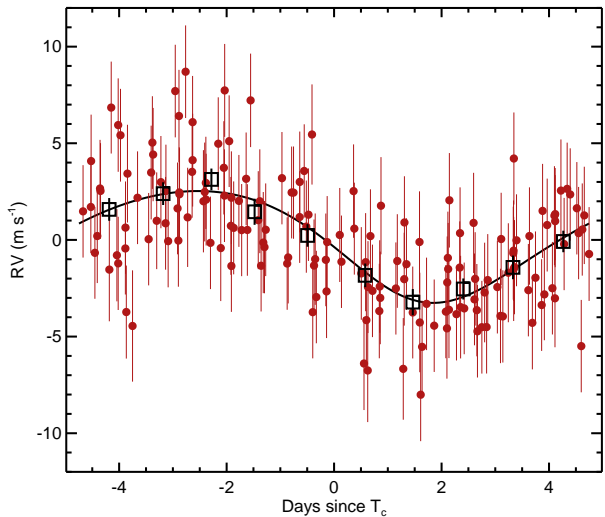}
\caption{RV measurements (red filled circles) phased to the best-fit orbital period and phase of the Keplerian model (see Section 3). The best-fit model is overplotted as a black line.  
RV measurements are binned in 0.1 phase intervals (black open squares) and have an RMS to the model of 0.36 m\,s$^{-1}$.  }
\label{fig:rv_phased}
\end{center}
\end{figure}

The orbital solution derived from modeling the RVs in Table \ref{tab:rvs} is consistent with the single planet models reported in \cite{How11} and H11.  
A Lomb-Scargle periodogram (Figure \ref{fig:running_periodogram}, top panel) of the RVs shows a prominent peak at 9.493 days, consistent with those models. We fit the RVs with the partially linearized, least-squares fitting procedure described in \cite{Wri09}, giving the best-fit solution shown in Figure \ref{fig:rv_phased}. 

To quantify the best-fit orbital solution, we used a Markov Chain Monte Carlo (MCMC) algorithm \citep{For05,For06} and report the median, 84.1\%, and 15.9\% levels of the marginalized posterior distributions. The likelihood was taken to be $e^{-\chi^2/2}$, where $\chi^2$ is the usual sum of the standardized residuals between the observed and calculated RVs. We adopted a Gregory eccentricity prior to correct for noise bias \citep{Gre10} and non-informative priors on other parameters. We excluded three outliers that were $\geq$7 m\,s$^{-1}$ from the best-fit model that has a residual RMS of 2.44 m\,s$^{-1}$ with the remaining measurements. These outliers are included in Table \ref{tab:rvs} and their timestamps in BJD-2,440,00 are 14928.963, 15529.170 and 15556.136. The MCMC analysis gives an orbital period $P = 9.4930 \pm 0.0021$ days, predicted time of mid transit $T_c$ = 2,455,982.59 $\pm$ 0.31 (BJD; \cite{East10}), eccentricity $e = 0.13 ^{+0.09}_{-0.06}$, longitude of pericenter $\omega = 120 ^{+95}_{-67}$ deg, velocity semi-amplitude $K = 2.86 \pm 0.27$ m\,s$^{-1}$, and semi-major axis $a = 0.0797 \pm 0.0007$ AU. When combined with the stellar mass of $0.75 \pm 0.02$ M$_{\odot}$ \citep{Hen11}, the minimum planet mass is  $M_p\sin i =  7.7 \pm 0.7$ $M_\earth$. Our estimates of $e \cos \omega = -0.02 \pm 0.10$ and $e \sin \omega = +0.07 \pm 0.09$ suggest that the planet's eccentricity is consistent with zero. Indeed, the MCMC analysis excludes $e > 0.26$ with 95\% confidence. The ephemeris based on this analysis (and not including any timing constraints from photometry) is more precise than, but consistent with, the ephemeris reported in \cite{How11}. 

The new RVs strengthen the evidence for a planet with an orbital period of $\approx$9.5 days, as shown visually in the bottom panel of Figure \ref{fig:running_periodogram}. The plot shows the Lomb-Scargle periodogram power \citep{Lom76,Sca82} at the planet's orbital period rising nearly monotonically as additional measurements are added. False periodicities, such as those due to spots, are typically only briefly coherent and may reappear with slightly different periods.

\section{Discussion}

We do not find a transit of the planet HD 97658b with the parameters obtained by H11 in the MOST photometry. Our observations further allow us to rule out transits for planetary radii as small as 1.87 R$_{\oplus}$, corresponding to a density less than 6.92 g cm$^{-3}$, within the predicted 3$\sigma$ transit window. Theoretically, it is still possible that the planet transits but is smaller and thus denser than these limits. For example, Kepler-10b \citep{Bat11} and Corot-7b \citep{Leg09,Hat11} are two known super-Earth exoplanets with densities above 7.0 g cm$^{-3}$. However, the parameters of such a transit would not match those reported by H11 and thus it could not account for the transit-like feature they observed.

We consider the possibility that the planet could have been previously transiting, but dynamical perturbations have caused its trajectory to drift off the disk of the star. However, this is very unlikely to have occurred after only 29 orbital cycles (since the last transit feature reported based on the APT photometry), especially since there is no sign of an additional companion in the radial velocities. We note that the peak at around 7 days in the top panel of Figure \ref{fig:running_periodogram} is a monthly alias of HD 97658b.

G. Henry made new APT observations of HD 97658 during the current 2012 observing season in an attempt to obtain improved light curves of the transit. On five separate nights between 18 January and 17 March, he was unable to find transits at the times predicted by H11. Therefore, he withdrew the paper stating that ``Additional observations are required to confirm or exclude transits of HD 97658b". The 2011 APT observations were acquired well past opposition of HD 97658 and so were made at relatively high air mass (up to 2.0) and differential air mass (up to 0.15). Thus, small changes in extinction or small errors in the applied extinction coefficients (as small as 0.01 mag/air mass) could result in systematic errors in the observed differential magnitudes of 0.001 mag or so, matching the claimed depth of the transits.

We emphasize that the MOST photometry allows us to exclude transits to high confidence given the mid-transit time and uncertainties predicted from the H11 model based on both RV measurements and APT photometry. If we consider the ephemeris derived from the RV measurements alone, we can rule out transits only between 0.55 and 3.6$\sigma$ of the predicted transit center, for planets with radii larger than 1.87 R$_{\oplus}$. 

The results presented in this Letter highlight the importance of independent follow-up observations, especially for a discovery as notable as that of a transiting super-Earth.

\section{Acknowledgments}

We thank Heather Knutson, Peter McCullough and David Anderson for helpful conversations, as well as the many observers who contributed to the measurements reported here. We are thankful to the referee, Dan Fabrycky, for his comments and suggestions which have significantly improved this manuscript. We gratefully acknowledge the efforts and dedication of the Keck Observatory staff, especially Scott Dahm, Hien Tran, and Grant Hill for support of HIRES and Greg Wirth for support of remote observing. We extend special thanks to those of Hawai`ian ancestry on whose sacred mountain of Mauna Kea we are privileged to be guests. Without their generous hospitality, the Keck observations presented herein would not have been possible.

DD is supported by a University of British Columbia Four Year Fellowship. Funding for the Stellar Astrophysics Centre (SAC) is provided by The Danish National Research Foundation and research is supported by the ASTERISK project (ASTERoseismic Investigations with SONG and Kepler) funded by the European Research Council (Grant agreement no.: 267864). The Natural Sciences and Engineering Research Council of Canada supports the research of DBG, JMM, AFJM and SMR. Additional support for AFJM comes from FQRNT (Qu\'ebec). RK and WWW were supported by the Austrian Science Fund (P22691-N16) and by the Austrian Research Promotion Agency-ALR.

\begin{deluxetable}{lccc}
\tabletypesize{\footnotesize}
\tablecaption{Radial Velocities and $S_{\mathrm{HK}}$ values for HD\,97658
\label{tab:rvs}}
\tablewidth{0pt}
\tablehead{
\colhead{}         & \colhead{Radial Velocity}     & \colhead{Uncertainty}  & \colhead{}  \\
\colhead{BJD -- 2,440,000}   & \colhead{(m\,s$^{-1}$)}  & \colhead{(m\,s$^{-1}$)}  & \colhead{$S_\mathrm{HK}$}
}
\startdata
 13398.04020 &    7.64 &    0.65  &         0.1970                     \\ 
 13748.03542 &    4.88 &    0.72  &         0.1900                     \\ 
 13806.96146 &    2.67 &    0.71  &         0.1870                     \\ 
 14085.15884 &   -4.74 &    0.80  &         0.1785                     \\ 
 14246.87816 &   -2.15 &    0.73  &         0.1760                     \\ 
 14247.83980 &   -4.67 &    0.94  &         0.1750                     \\ 
 14248.94470 &   -2.76 &    1.08  &         0.1690                     \\ 
 14249.80244 &    0.19 &    1.09  &         0.1740                     \\ 
 14250.83983 &    1.60 &    0.98  &         0.1740                     \\ 
 14251.89455 &   -0.07 &    0.95  &         0.1720                     \\ 
 14255.87144 &   -0.86 &    0.73  &         0.1735                     \\ 
 14277.81740 &   -0.86 &    0.99  &         0.1770                     \\ 
 14278.83838 &    0.21 &    0.99  &         0.1750                     \\ 
 14279.83000 &    2.37 &    1.01  &         0.1760                     \\ 
 14294.76351 &   -5.33 &    1.11  &         0.1690                     \\ 
 14300.74175 &   -0.33 &    1.02  &         0.1720                     \\ 
 14304.76223 &   -5.64 &    1.17  &         0.1740                     \\ 
 14305.75803 &   -4.99 &    0.80  &         0.1740                     \\ 
 14306.77175 &   -4.08 &    1.02  &         0.1690                     \\ 
 14307.74766 &   -0.86 &    0.78  &         0.1750                     \\ 
 14308.75036 &    2.43 &    0.75  &         0.1755                     \\ 
 14309.74773 &    1.42 &    1.07  &         0.1760                     \\ 
 14310.74343 &   -0.04 &    1.03  &         0.1750                     \\ 
 14311.74391 &    4.39 &    1.14  &         0.1760                     \\ 
 14312.74242 &   -2.21 &    1.03  &         0.1770                     \\ 
 14313.74419 &   -1.17 &    1.21  &         0.1780                     \\ 
 14314.75074 &   -0.18 &    1.14  &         0.1740                     \\ 
 14455.15432 &   -5.21 &    1.11  &         0.1820                     \\ 
 14635.79759 &    0.71 &    0.98  &         0.1750                     \\ 
 14780.12544 &   -3.78 &    1.12  &         0.1770                     \\ 
 14807.09051 &   -7.73 &    1.22  &         0.1730                     \\ 
 14808.15781 &   -4.54 &    1.27  &         0.1710                     \\ 
 14809.14349 &   -1.62 &    1.16  &         0.1730                     \\ 
 14810.02507 &    1.49 &    1.25  &         0.1730                     \\ 
 14811.11469 &   -2.60 &    1.31  &         0.1730                     \\ 
 14847.11818 &   -2.56 &    1.34  &         0.1720                     \\ 
 14927.89832 &   -1.20 &    1.19  &         0.1700                     \\ 
 14928.96319 &   -9.26 &    1.13  &         0.1700                     \\ 
 14929.84171 &   -7.82 &    1.30  &         0.1690                     \\ 
 14954.97010 &    0.11 &    1.10  &         0.1710                     \\ 
 14955.92354 &    0.96 &    0.59  &         0.1723                     \\ 
 14956.90526 &    1.39 &    0.59  &         0.1717                     \\ 
 14963.96620 &    1.93 &    0.62  &         0.1687                     \\ 
 14983.87422 &   -1.21 &    0.65  &         0.1703                     \\ 
 14984.90305 &   -0.54 &    0.66  &         0.1710                     \\ 
 14985.84495 &   -4.02 &    0.65  &         0.1710                     \\ 
 14986.88981 &   -3.70 &    0.65  &         0.1697                     \\ 
 14987.89641 &   -4.37 &    0.64  &         0.1697                     \\ 
 14988.84295 &   -5.79 &    0.65  &         0.1703                     \\ 
 15041.75244 &    0.94 &    1.33  &         0.1690                     \\ 
 15164.11579 &    0.94 &    1.25  &         0.1730                     \\ 
 15188.15834 &   -3.08 &    0.73  &         0.1703                     \\ 
 15190.13345 &   -6.56 &    0.64  &         0.1703                     \\ 
 15191.15996 &   -4.79 &    0.69  &         0.1700                     \\ 
 15192.12876 &   -1.09 &    0.63  &         0.1713                     \\ 
 15193.11580 &    1.12 &    0.66  &         0.1720                     \\ 
 15197.14377 &   -1.99 &    0.66  &         0.1713                     \\ 
 15198.06458 &   -3.50 &    0.67  &         0.1720                     \\ 
 15199.08947 &   -3.56 &    0.65  &         0.1723                     \\ 
 15256.95734 &    3.02 &    0.68  &         0.1800                     \\ 
 15285.94174 &   -2.27 &    0.71  &         0.1750                     \\ 
 15289.83018 &   -1.17 &    0.61  &         0.1780                     \\ 
 15311.78391 &   -5.57 &    0.68  &         0.1730                     \\ 
 15312.85871 &   -3.88 &    0.57  &         0.1727                     \\ 
 15313.76705 &    0.42 &    0.64  &         0.1723                     \\ 
 15314.78098 &    1.12 &    0.67  &         0.1723                     \\ 
 15317.96171 &    0.24 &    0.65  &         0.1740                     \\ 
 15318.94529 &   -2.79 &    0.66  &         0.1747                     \\ 
 15319.90227 &   -4.81 &    0.59  &         0.1763                     \\ 
 15320.86046 &   -4.60 &    0.58  &         0.1800                     \\ 
 15321.83394 &   -1.08 &    0.59  &         0.1807                     \\ 
 15342.87709 &   -1.85 &    0.62  &         0.1757                     \\ 
 15343.83030 &   -1.12 &    0.64  &         0.1757                     \\ 
 15344.87915 &    1.23 &    0.66  &         0.1750                     \\ 
 15350.78148 &   -4.43 &    0.58  &         0.1730                     \\ 
 15351.88523 &    0.64 &    0.60  &         0.1740                     \\ 
 15372.75567 &    2.46 &    0.59  &         0.1790                     \\ 
 15373.78450 &   -0.55 &    0.57  &         0.1787                     \\ 
 15374.75746 &    0.12 &    0.58  &         0.1777                     \\ 
 15375.77510 &   -0.48 &    0.59  &         0.1767                     \\ 
 15376.74459 &   -2.32 &    0.57  &         0.1773                     \\ 
 15377.74203 &   -0.71 &    0.56  &         0.1773                     \\ 
 15378.74358 &    3.15 &    0.61  &         0.1760                     \\ 
 15379.79184 &    1.29 &    0.63  &         0.1763                     \\ 
 15380.74383 &    5.79 &    0.60  &         0.1753                     \\ 
 15400.74199 &   -0.20 &    0.67  &         0.1767                     \\ 
 15401.76937 &   -1.48 &    1.43  &         0.1810                     \\ 
 15403.73759 &   -3.72 &    0.72  &         0.1757                     \\ 
 15404.73679 &   -4.06 &    0.65  &         0.1813                     \\ 
 15405.74070 &   -5.50 &    0.65  &         0.1810                     \\ 
 15406.73719 &   -3.14 &    0.60  &         0.1817                     \\ 
 15407.75814 &    0.44 &    0.77  &         0.1800                     \\ 
 15410.73844 &    5.03 &    0.65  &         0.1790                     \\ 
 15411.73481 &    2.10 &    0.67  &         0.1780                     \\ 
 15412.73197 &    1.94 &    1.18  &         0.1780                     \\ 
 15413.73361 &    1.46 &    0.72  &         0.1633                     \\ 
 15501.14886 &   -2.49 &    0.64  &         0.1827                     \\ 
 15522.13389 &    1.58 &    0.64  &         0.1820                     \\ 
 15529.17006 &  -12.59 &    1.56  &         0.1770                     \\ 
 15543.14746 &    1.47 &    0.65  &         0.1847                     \\ 
 15546.12332 &   -2.09 &    0.66  &         0.1827                     \\ 
 15556.13576 &   12.54 &    0.71  &         0.1843                     \\ 
 15557.07525 &   -3.09 &    0.68  &         0.1860                     \\ 
 15559.12764 &   -2.47 &    0.78  &         \nodata                   \\ 
 15585.09918 &   -3.47 &    0.65  &         0.1810                     \\ 
 15605.98568 &   -5.58 &    0.67  &         0.1860                     \\ 
 15606.98402 &   -3.02 &    0.66  &         0.1863                     \\ 
 15607.98123 &   -1.79 &    0.75  &         0.1847                     \\ 
 15614.03861 &   -0.16 &    0.64  &         0.1897                     \\ 
 15614.87512 &    0.99 &    0.81  &         0.1890                     \\ 
 15633.99315 &   -4.90 &    0.66  &         0.1887                     \\ 
 15635.05051 &   -1.75 &    0.69  &         0.1883                     \\ 
 15635.99669 &   -1.28 &    0.63  &         0.1883                     \\ 
 15636.75731 &   -1.73 &    0.59  &         0.1870                     \\ 
 15663.88480 &   -5.35 &    0.69  &         0.1870                     \\ 
 15667.96860 &   -0.55 &    0.61  &         0.1893                     \\ 
 15668.93400 &    1.39 &    0.63  &         0.1893                     \\ 
 15670.83890 &   -3.57 &    0.61  &         0.1887                     \\ 
 15671.81197 &   -2.57 &    0.63  &         0.1890                     \\ 
 15672.79841 &   -1.01 &    0.65  &         0.1887                     \\ 
 15673.80273 &   -0.09 &    0.61  &         0.1873                     \\ 
 15696.87321 &   -1.43 &    0.64  &         0.1880                     \\ 
 15697.79492 &   -2.39 &    0.61  &         0.1877                     \\ 
 15698.79967 &   -3.50 &    0.59  &         0.1883                     \\ 
 15699.80627 &   -6.59 &    0.62  &         0.1883                     \\ 
 15700.82486 &   -4.69 &    0.63  &         0.1887                     \\ 
 15703.77767 &   -1.50 &    0.61  &         0.1873                     \\ 
 15704.74908 &    0.57 &    0.59  &         0.1900                     \\ 
 15705.75018 &   -2.41 &    0.59  &         0.1903                     \\ 
 15706.80704 &   -1.97 &    0.59  &         0.1910                     \\ 
 15707.79932 &   -2.19 &    0.63  &         0.1920                     \\ 
 15723.76737 &    1.40 &    0.43  &         0.1862                     \\ 
 15728.75382 &   -3.26 &    0.60  &         0.1860                     \\ 
 15731.79487 &    1.47 &    0.88  &         0.1897                     \\ 
 15733.75978 &    1.02 &    0.59  &         0.1890                     \\ 
 15734.78290 &   -2.40 &    0.55  &         0.1893                     \\ 
 15735.78926 &   -2.06 &    0.54  &         0.1893                     \\ 
 15738.75421 &   -4.14 &    0.58  &         0.1877                     \\ 
 15751.74696 &    3.97 &    0.66  &         0.1967                     \\ 
 15752.74130 &    1.89 &    0.62  &         0.1970                     \\ 
 15759.74698 &   -0.52 &    0.62  &         0.1887                     \\ 
 15760.73902 &   -0.42 &    0.63  &         0.1913                     \\ 
 15761.74431 &    1.30 &    0.60  &         0.1900                     \\ 
 15762.75287 &   -0.44 &    0.69  &         0.1883                     \\ 
 15763.74854 &   -2.28 &    0.65  &         0.1910                     \\ 
 15768.73583 &    0.27 &    0.62  &         0.1883                     \\ 
 15770.74725 &    3.35 &    0.71  &         0.1810                     \\ 
 15871.12714 &   -4.77 &    0.71  &         0.1830                     \\ 
 15878.14302 &   -4.80 &    0.64  &         0.1927                     \\ 
 15879.09674 &   -7.46 &    0.70  &         0.1923                     \\ 
 15880.15181 &   -9.07 &    0.68  &         0.1907                     \\ 
 15882.15606 &   -3.66 &    0.64  &         0.1903                     \\ 
 15902.04440 &    0.58 &    0.64  &         0.1940                     \\ 
 15903.04465 &    4.36 &    0.67  &         0.1933                     \\ 
 15904.13372 &    5.35 &    0.63  &         0.1950                     \\ 
 15905.06744 &    4.05 &    0.57  &         0.1950                     \\ 
 15929.14692 &   -5.02 &    0.62  &         0.1987                     \\ 
 15932.04835 &   -1.02 &    0.64  &         0.1957                     \\ 
 15945.09826 &   -0.81 &    0.77  &         0.2053                     \\ 
 15961.02344 &    6.64 &    0.66  &         0.2050                     \\ 
 15967.94465 &   -0.29 &    0.68  &         0.1963                     \\ 
 15972.95568 &   -0.37 &    0.62  &         0.2013                     \\ 
 15990.90524 &    6.16 &    0.73  &         0.2030                     \\ 
 15991.90284 &    2.51 &    0.71  &         0.2030                     \\ 
 15997.00945 &   -0.71 &    0.63  &         0.2000                     \\ 
 15999.79702 &    3.92 &    0.68  &         0.2010                     \\ 
 16018.90330 &    6.67 &    0.68  &         0.2040                     \\ 
 16019.96914 &    2.13 &    0.63  &         0.2047                     \\ 
 16027.79949 &    3.07 &    0.67  &         0.2077                     
\enddata
\end{deluxetable}

\end{document}